\documentclass{aa}     
\usepackage{natbib}
\usepackage{graphicx}
\usepackage[mathcal]{eucal}
\usepackage{amssymb} 
\usepackage{amsmath}
\usepackage[T1]{fontenc}
\usepackage{txfonts}

\def\ds {{$\delta$~Scuti~}}
\def\dss {{$\delta$~Scuti~stars}}

\def\teff {{T_{\mathrm{eff}}}}

\def\Oi {{\Omega_{\mathrm{i}}}}

\def\muHz {{\mu\mbox{Hz}}}
\def\kms {{\mathrm{km}\,\mathrm{s}^{-1}}}
\def\msol {{\mathrm{M}_\odot}}

\def\ratio {{\Pi_{1/0}}}
\def\ratiorot {{\Pi_{1/0}\,(\Omega)}}
\def\ratiorotdeg {{\Pi_{1/0}^{\mathrm{d}}\,(\Omega)}}
\def\dratiod {\delta\ratio^{\mathrm{d}}}

\def\paperI {{Paper~I}}

\def\dg {{DG92}}
\def\soufi {{SGD98}}
\def\cesam {{\sc cesam}}
\def\filou {{\sc filou}}

\def\bs {\boldsymbol}

\def\msol {{\mathrm{M}_\odot}}

\def\i {{\emph{i}}}

\def\wdt {{\tilde{\omega}_2}}

\def\xig   {{\bs {\xi}}}

\newcommand{\eqn} [1] {
\begin{equation}
#1
\end{equation}}

\begin{document}

   \title{The role of rotation on Petersen Diagrams}
   \subtitle{II. The influence of near-degeneracy}

   \titlerunning{The influence of near-degeneracy on rotating Petersen Diagrams}
   \authorrunning{Su\'arez, et al.}

   \author{J.C. Su\'arez\inst{1,2}
   \thanks{Associate researcher at institute (2).}
   \and R. Garrido\inst{1}  
   \and A. Moya\inst{1}}

   \offprints{J.C Su\'arez~\email{jcsuarez@iaa.es}}

   \institute{Instituto de Astrof\'{\i}sica de Andaluc\'{\i}a (CSIC), CP3004, Granada, Spain 
	      \and 
	      Observatoire de Paris, LESIA, UMR 8109, Meudon, France}

   \date{Received ... / Accepted ...}

   \abstract
            {}
	     {In the present work, the effect of near-degeneracy on 
rotational Petersen 
	     diagrams (RPD) is analysed.}
	     {Seismic models are computed considering rotation effects on
              both equilibrium models and adiabatic oscillation frequencies 
	      (including second-order near-degeneracy effects). Contamination
	      of coupled modes and coupling strength on the first radial modes
	      are studied in detail.}
             {Analysis of relative intrinsic amplitudes of near-degenerate modes
              reveals that the identity of the fundamental radial mode and its 
	      coupled quadrupole pair are almost unaltered once near-degeneracy 
	      effects are considered. However, for the first overtone, a mixed
	     radial/quadrupole identity is always predicted. The effect of 
             near-degeneracy on the oscillation frequencies becomes critical for 
	     rotational velocities larger than $15-20\,\kms$, for which large 
	     wriggles in the evolution of the period ratios are obtained 
	     (up $10^{-2}$). Such wriggles imply uncertainties, in terms of metallicity 
	     determinations using RPD, reaching up to 0.50\,dex, 
	     which can be critical for Pop.~I HADS (High Amplitude \dss). In terms of mass 
	     determinations, 
	     uncertainties reaching up to $0.5\,\msol$ are predicted. The location 
	     of such wriggles is found to be independent of metallicity and rotational 
	     velocity, and governed mainly by the avoided-crossing phenomenon. 
	     }   
	     {Near-degeneracy affects significantly the $\ratiorot$ period
              ratios even for relatively low rotational velocities, and that can be 
	      critical when accurate determinations of mass and metallicity are required.
              Nevertheless, analysis of near-degeneracy effects provides some clues
	     for the identification of the fundamental radial mode, the first overtone, 
	     and their corresponding quadrupole coupled pairs.
	     This can be especially useful when additional information on mode
             identification and/or metallicity is available, for example from multicolour 
	     photometry and/or spectroscopy, not only for accurate diagnostics on metallicity 
	     and mass, but also because it is possible to constrain, to some extent, 
	     the rotational velocity of the star (and thereby its inclination angle). 
	     } 
	     

\maketitle


\section{Introduction\label{sec:intro}}


As explained in several papers \citep{DG92,Soufi98,Sua06rotcelSGM06}, the effects of
rotation on oscillations (near-degeneracy included) cannot be neglected for 
asteroseismic studies even for relatively slow-rotating stars. Near-degeneracy 
affects the small separations since it occurs for close modes (under certain 
selection rules, $\Delta\ell=0,\pm2$, and $\Delta\,m=0$). 
The effect of near-degeneracy on the oscillation modes is two-fold: it changes
the oscillation frequencies (through the coupling strength) and the mode 
character (mode contamination).
While the former effect is proportional to the rotation rate, the mode 
contamination is nearly independent of the rotation rate and strongly dependent 
on the mode degree $\ell$.

In the present paper we investigate the impact of including near-degeneracy in 
the asteroseismic diagnostics using rotational Petersen diagrams (hereafter RPD)
for the range of mass, metallicity and rotational velocities studied in 
\citet{Sua06pdrotPaperI} (from now on \paperI). In \paperI, the effect of rotation 
on period ratios of radial modes was explored focusing on double-mode pulsators. 
Detailed seismic models were computed considering rotation effects on both 
equilibrium models and on adiabatic oscillation frequencies. In particular, 
for $1.8\,\msol$ stellar models, period ratios were calculated for different 
rotational velocities (RPD) and metallicities and then compared with standard 
non-rotating Petersen diagrams (PD). We showed that the difference in period 
ratios increases with the rotational velocity for a given metallicity. Such
differences in period ratios were found equivalent to differences in the
metallicity of models up to $0.30\,\mathrm{dex}$. The detailed investigation of 
this \emph{confusion} scenario is therefore particularly relevant when accurate 
metallicity and/or mass determinations are required. As stated in \paperI, such 
ambiguity may still be increased when including different stellar masses, 
rotational velocities and metallicities, etc. Important effects are expected to 
be found when near-degeneracy is taken into account \citep{Alosha03}. In that 
work, the behaviour of $\ratiorot$ when near-degeneracy effects are included  
were studied for a typical $1.8\,\msol$ \ds\ stellar model with a surface
rotational velocity of $100\,\kms$. He showed that very large and non-regular 
perturbations of such ratios are expected to occur.


The paper is structured as follows: General description of the modelling,
focusing on how rotation is taken into account is given in 
Sect.~\ref{sec:modelling}; in Sect.~\ref{sec:neardeg} the effect of 
near-degeneracy on the oscillation frequencies, as well as the
coupled mode contamination, is analysed for the fundamental radial mode 
and the first overtone; in Sect.~\ref{sec:impact_Z}, the impact of 
near-degeneracy on the RPD diagnostic diagrams is discussed; finally, 
conclusions are given in Sect.~\ref{sec:conclusions}.

  
\section{The modelling\label{sec:modelling}}   

Proceeding as in \paperI, equilibrium models are computed with the
evolutionary code \cesam\ \citep{Morel97}, for which a first-order effect of
rotation is taken into account in equilibrium equations. Uniform rotation and 
global conservation of the total angular momentum is assumed. We recall that 
this first-order effects of rotation are considered by subtracting the 
spherically averaged contribution of the centrifugal acceleration to the gravity 
of the model. This spherically averaged component of the centrifugal acceleration 
does not change the order of the hydrostatic equilibrium equations. Such models 
are the so-called "pseudo-rotating" models \citep[see][]{Soufi98, Sua06rotcelSGM06}. 
Although the non-spherical components of the centrifugal acceleration are not 
considered, they are included as a perturbation in the oscillation frequencies 
computation.

A grid of equilibrium models is built, similar to the one used in \paperI, 
which is composed of models with masses ranging from 1.80 to $2\,\msol$, 
with seven different metallicities: 
$\mathrm{[Fe/H]}=0,-0.1,-0.2,-0.30,-0.35,-0.50$ and $-1.00\,\mathrm{dex}$.
The initial (ZAMS) rotational velocity of models is restricted 
to $\Oi=15-50\,\kms$. We recall that, due to the global conservation of the
total angular momentum during the evolution, the rotational velocity of the
models can decrease up to $0.75\,\Oi$ at the end of the main sequence (close to
the terminal age main sequence, TAMS). A mixing-length parameter of 
$\alpha_{ML}=1.80$ and an overshooting parameter of $d_{\mbox{ov}}=0.2$ are assumed. 

Computation of the oscillation spectra is carried out using the oscillation
code \filou\ \citep{filou,SuaThesis}, which is based on a perturbative analysis 
and provides adiabatic oscillations, corrected for the effects of rotation up 
to the second order (centrifugal and Coriolis forces), including
near-degeneracy effects.

    
\section{Near-degenerate oscillation frequencies\label{sec:neardeg}}

In order to analyse the role of rotation on $\ratiorot$ period ratios, the
complete second-order (in $\nu_\Omega$, the rotation rate) formalism given in
\citet{Sua06rotcelSGM06} (hereafter SGM06) is considered. 
The near-degeneracy effect can be divided into two: the 
effect on the oscillation frequencies, and the mode contamination, i.e.
the weight of the original individual spherical harmonics describing 
the oscillation mode in the resulting coupled mode. Both aspects
are theoretically analysed in \citet{Lochard07} for solar-like, main-sequence
stars. In that work, which is based on the developments given in SGM06, 
the authors use a method to numerically remove such \emph{disturbing} 
effects on the small separations in order to analyse the echelle diagrams 
for rotating stars free of rotational effects from the observed frequencies.
In the present work, we apply the developments given in those works to
the specific case of the $\ratiorot$ period ratios. To do so, we 
selected from the model grid described in Sect.~\ref{sec:modelling}, 
eight models (see Table~\ref{tab:modelcarac}) with similar evolutionary stage
(1.3\,Gyr), mean density, effective temperature, and with two different initial 
rotational velocities: $\Omega_i=25$, and $50\,\kms$. The objectives of this
selection are twofold: first, we are interested in having models representative
of High Amplitude \dss\ (HADS), i.e. with evolutionary stages near the end of the 
main sequence, and second, the models should have similar global characteristics but for
different rotational velocities and metallicities (models $\mathrm{M}_{1i}$
and $\mathrm{M}_{2i}$, with $i=1,4$, of Table~\ref{tab:modelcarac},
respectively), which permits the study of both effects separately.
\begin{table}
  \begin{center}
    \caption{Characteristics of the eight selected $1.8\,\msol$ models 
             with similar evolutionary stages ($\sim1.3 Gyr$). From left
	     to right, $\Omega$ is the surface rotational velocity, 
	     $\teff$ the effective temperature (in logarithmic scale), 
	     $X_{\mathrm{c}}$ the central hydrogen fraction, $\bar{\rho}$, 
	     the mean density, and the last column is the metallicity. 
	     ${\rm M}_{1i}$ and ${\rm M}_{2i}$ models have rotational velocities
	     of $\Omega\sim20$ and $41\,\kms$, respectively.}
    \vspace{1em}
    \renewcommand{\arraystretch}{1.2}
    \begin{tabular}[ht]{cccccc}
      \hline
        Model & $\Omega$  & $\log\teff$ & $X_{\mathrm{c}}$ &  $\bar{\rho}$ & [Fe/H] \\
              &  ($\kms$)         & (K)  &  & ($\mathrm{g}\,\mathrm{cm}^{-3}$) & dex\\
      \hline
       ${\rm M}_{11}$   & 21.05 & 3.83 & 0.187 & 0.114 &  0.00\\
       ${\rm M}_{12}$   & 20.60 & 3.84 & 0.177 & 0.114 & -0.10\\
       ${\rm M}_{13}$   & 20.10 & 3.86 & 0.170 & 0.121 & -0.20\\
       ${\rm M}_{14}$   & 19.90 & 3.89 & 0.190 & 0.149 & -0.30\\
      \hline
       ${\rm M}_{21}$   & 42.25 & 3.83 & 0.188 & 0.113 &  0.00 \\
       ${\rm M}_{22}$   & 41.30 & 3.85 & 0.177 & 0.114 & -0.10 \\
       ${\rm M}_{23}$   & 40.20 & 3.86 & 0.170 & 0.121 & -0.20 \\
       ${\rm M}_{24}$   & 39.91 & 3.89 & 0.190 & 0.149 & -0.30 \\            
      \hline
      \end{tabular}
    \label{tab:modelcarac}
  \end{center}
\end{table}

\begin{figure*}[t]
 \begin{center}  
   \includegraphics[width=8.8cm]{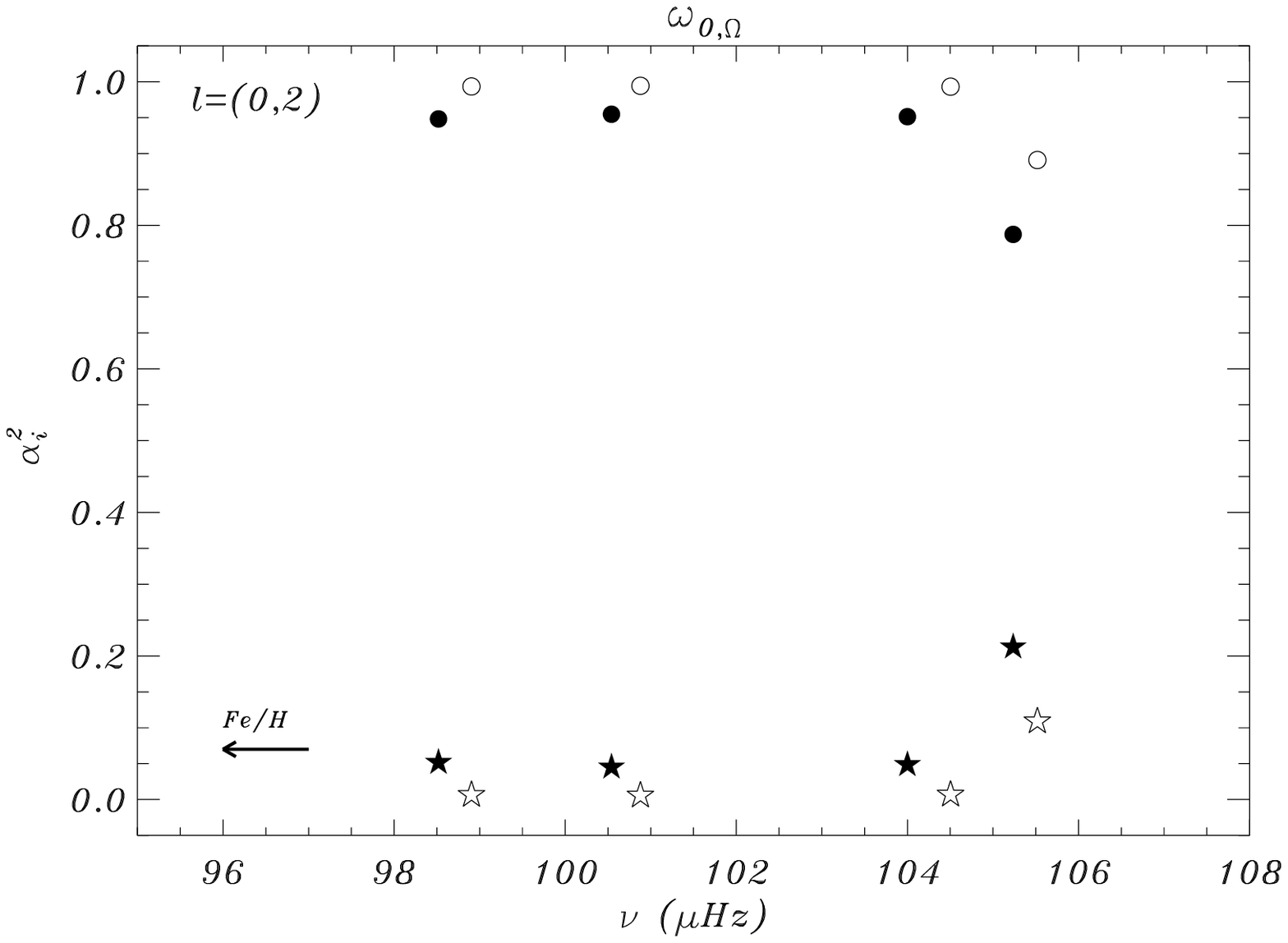}\hspace{-0.40cm}
   \includegraphics[width=8.8cm]{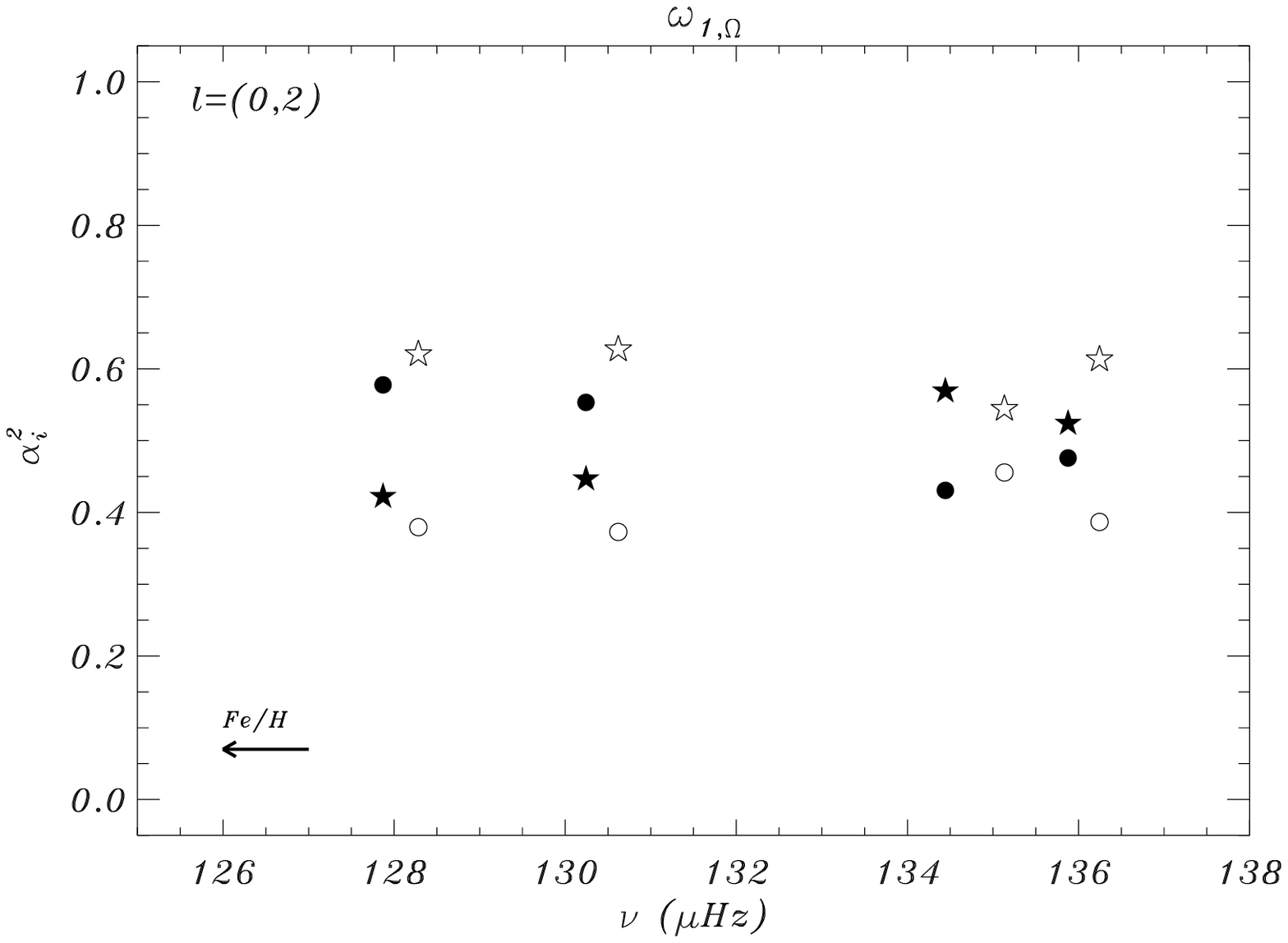}
   \caption{Contamination coefficients $\alpha_i^2$ of degenerate pairs 
            $\ell=(0,2)$ as a function of the oscillation frequency $\nu$
            (in $\muHz$) for the selected models listed in 
	    Table~\ref{tab:modelcarac}. Left and right panels show 
	    $\alpha_i^2$ for the fundamental radial mode $\omega_{0,\Omega}$ 
	    and the first overtone, respectively. As in Fig.~\ref{fig:Cab_p0p1}, 
	    filled symbols represent $\alpha_i^2$ calculated for the selected 
	    models with $\Omega\sim41\,\kms$, and empty symbols represent
	    $\alpha_i^2$ calculated for the selected models with $\Omega\sim20\,\kms$.
	    Circles and stars represent the radial modes and the corresponding
            $\ell=2, m=0$ modes, respectively.}
   \label{fig:contam}
 \end{center}
\end{figure*}

\subsection{Mode contamination \label{ssec:contam}}

Mode identification is implicit when performing analysis using PD. Indeed,
if no additional evidences points towards the observed frequencies as belonging to
the corresponding radial mode and the first overtone, the period $\ratio$ by
itself, does not guarantee such identification. In this context, when rotation 
is present, the mode contamination plays an important role, since it may change 
the intrinsic identity of the coupled modes.
 
It was shown by \citet{Pagoda02} that rotational coupling affects the diagnostic
diagrams for discriminating the angular degree $\ell$ of a particular mode 
using multicolour photometry. In particular, the authors found that
loci in the amplitude ratio--phase difference diagrams \citep{Garrido00} become
both aspect and $m$-dependent. In \citet{Casas06}, it was found that mode 
contamination is significant for radial modes in \dss. Therefore, mode
contamination should be then taken into account for the correct identification 
using amplitude ratio--phase difference diagnostic diagrams. We investigate here
how the fundamental radial mode and the first overtone are \emph{contaminated} by
their corresponding $\ell=2, m=0$ coupled pairs.

Coupled modes are described by a linear combination of the involved modes. 
Following SGM06, for two\footnote{The formalism is similar for three or more
modes.} degenerate modes, $j$ and $k$, the so called \emph{contamination coefficients}, 
$\alpha_i$ ($i=j,k$), defined in \citet{Lochard07}, are written as
\eqn{\alpha_{jk}=\frac{\alpha_j}{\alpha_k}= 
     \frac{m_{jk}^{(1)}+m_{jk}^{(2)}}{m_{jj}^{(1)}+m_{jj}^{(2)}
              \pm\delta\omega_{0}/2+
	      \delta\omega_{0}^2/8}\,,\label{eq:defstjk}}
where $m_{jk}^{(1)}$ and $m_{jk}^{(2)}$ are the elements of the first- and
second-order correction terms matrices ${\cal M}_{jk}$ 
(SGM06~Eqs.~22 and 23, respectively), and 
$\delta\omega_{0}=\omega_{0,j}-\omega_{0,k}$. The coefficients are constructed
such as the closing relation (normalisation)
\eqn{\alpha_{j}^2+\alpha_{k}^2=1\,.\label{eq:Sjk_norm}}
is imposed. The coefficients $\alpha_i$ are the relative amplitudes
of the degenerate modes, and can be interpreted as the relative weight of each
mode in $\xig$, or otherwise, the degree of \emph{contamination} of one
degenerate mode into another. 

For the selected models in Table~\ref{tab:modelcarac} we computed the relative 
intrinsic mode amplitudes $\alpha_i^2$ of the fundamental radial mode 
$\omega_{0,\Omega}$ , the first overtone $\omega_{1,\Omega}$, and their 
corresponding coupled pairs. The results are shown in Fig.~\ref{fig:contam}, 
in which the coefficients $\alpha_i^2$ are depicted as a function of the 
oscillation frequency of the fundamental radial mode. Similar $\alpha_i^2$ 
values are found for the two rotational velocities which 
confirms the independence of mode contamination upon
the rotational velocity. In the case of the fundamental radial mode
(left panel), we found that $\alpha_i^2\sim1$ for the radial modes and 
$\alpha_i^2\sim0$ for the coupled $\ell=2,m=0$ pairs, except for the
lowest metallic model, for which these values are 
$\alpha_j\sim0.8$ and $\alpha_k\sim0.2$, respectively. 
This means that both the fundamental radial mode and its coupled pair keep
their original identity and can be still regarded as a radial and a quadrupole
mode, respectively. On the contrary, the contamination coefficients for 
the first overtone and its coupled pair (right panel) are of the order of 
0.60 and 0.40, respectively, which implies that these rotationally coupled
modes exhibit a \emph{mixed} radial/quadrupole identity.  Notice that, for 
the model with $\Omega\sim41\,\kms$ (filled symbols), $\alpha_i^2\sim0.55-0.6$ for
the radial modes and $\alpha_i^2\sim0.40-0.45$ for their
coupled $\ell=2$ pairs. This situation is inverted for the
models with $\Omega\sim20\,\kms$ (empty symbols). Interestingly,
for the model with [Fe/H]=-0.20 ($\nu\sim134\,\muHz)$ and 
a rotational velocity of $\Omega\sim41\,\kms$, the coefficients are
re-inverted. The origin of such permutations is not evident, and
is probably related to the other aspect of these results, which
are not yet well understood: the different
behaviour of the $\alpha_i^2$ found for the fundamental radial mode
and the first radial overtone. Physically, the origin of such a different 
behaviour probably resides in the nature of their corresponding coupled pairs.
Indeed, the radial order of the quadrupole modes coupled with the
fundamental radial modes is systematically lower (in the range of
$g$ modes) than the radial order of the quadrupole modes coupled
with the first radial overtones. This enhances the weight
of the vertical and horizontal displacement eigenfunctions of the 
quadrupole modes in the coupling
terms $m_{jk}^{(1)}$ and $m_{jk}^{(2)}$ (Eq.~\ref{eq:defstjk}), near the stellar core and
surface, respectively. In order to better understand this, a work is
currently in progress to analyse in detail the complex dependence of 
$\alpha_i$ coefficients upon the eigenfunctions of the concerned 
coupled modes, together with variations in the structure of the star 
due to rotation and evolution.

This result is important, not only because it implies the existence of patterns in 
the contamination coefficients that may help us to identify properly the affected
modes, but also because it may give us relevant information about
the nature of the modes. It would be particularly interesting to compare
these results with those of the non-perturbative theory \citep{Reese06}.
Together with accurate multicolour photometry, analysis of mode contamination 
is such a powerful asteroseismic tool because it may help to discriminate,
not only the fundamental radial mode from the first overtone, but also to
identify the quadrupole modes coupled with them (if they are present
in the observed spectrum).

    
\subsection{Near-degeneracy effects on oscillation
frequencies\label{ssec:coupstrength}}

In addition to the problem of mode identification, attention is necessarily
focused on the effect of near-degeneracy on the oscillation frequencies. As
shown by \citet{Soufi98} and SGM06, such an effect cannot be neglected for asteroseismic
studies of moderately rotating stars. As is shown later on in the paper, although
HADS are relatively slow-rotating stars, and thereby near-degeneracy effects
on the frequencies are expected to be small, they can modify the $\ratiorot$
period ratios sufficiently to cause significant variations, even critical, 
in the PD. Such an effect is far from trivial and
deserves special attention.
\begin{figure*}[t]
 \begin{center}
   \includegraphics[width=9cm]{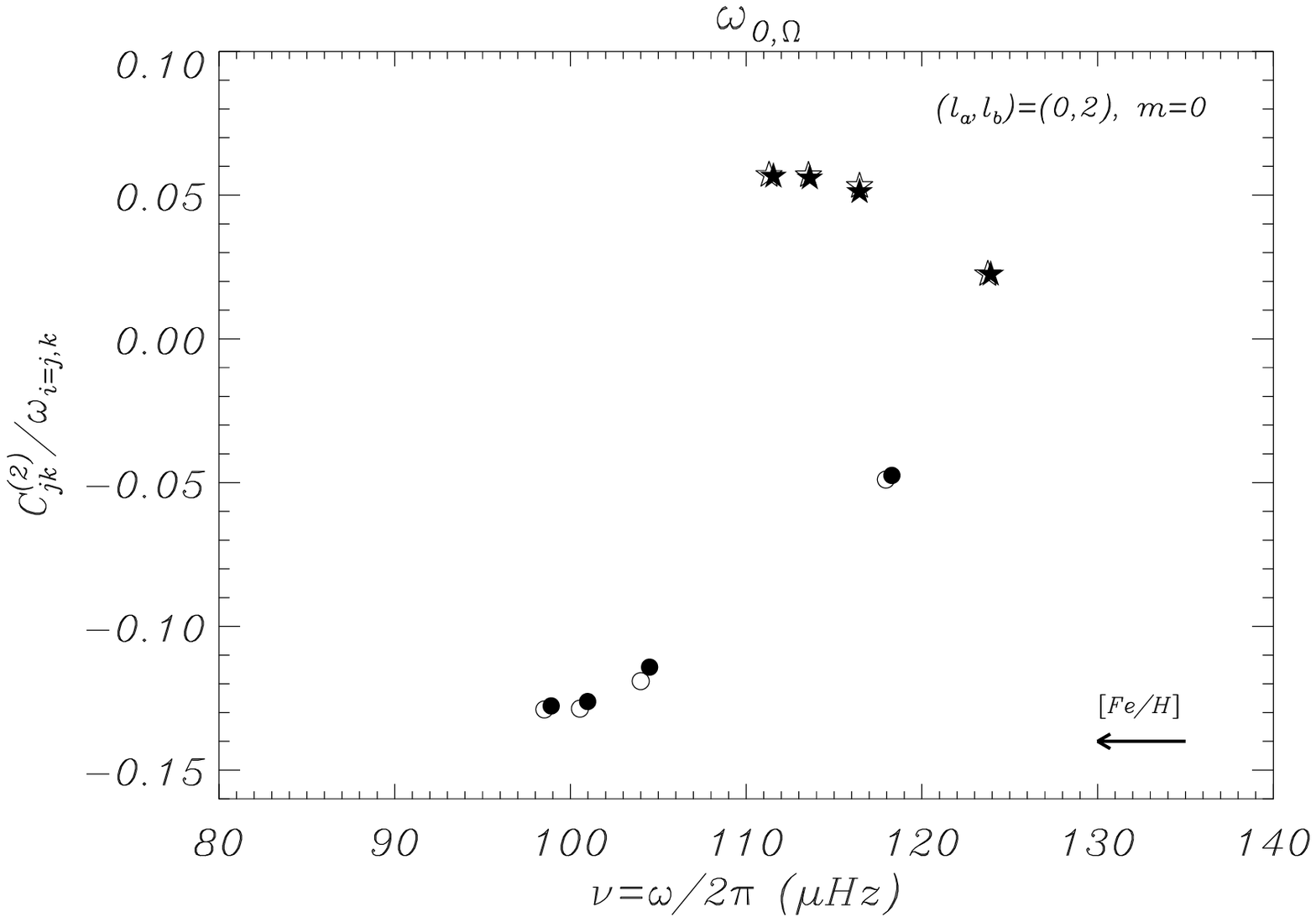}\hspace{-0.40cm}  
   \includegraphics[width=9cm]{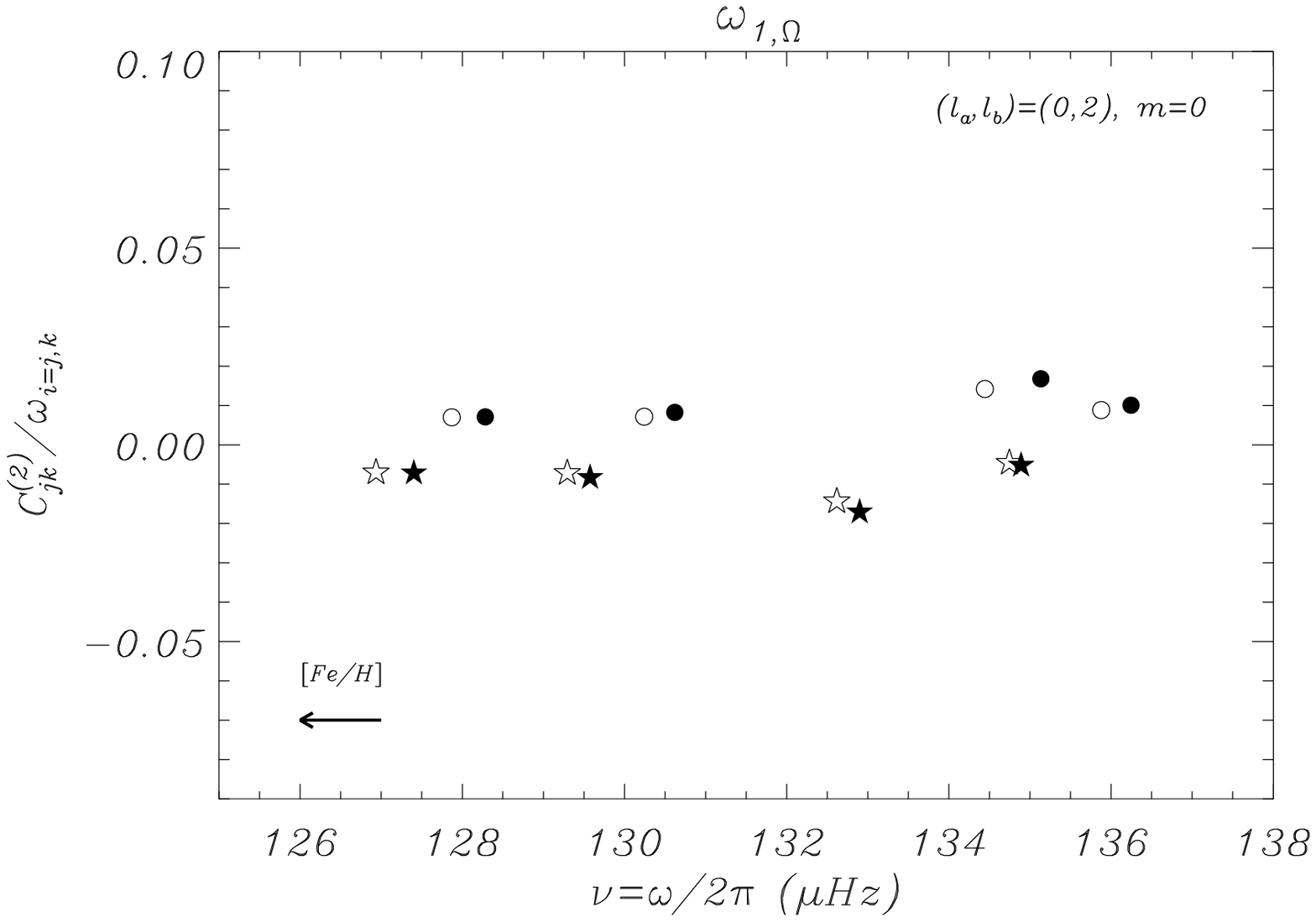}         
   \caption{Coupling strength coefficient normalised to the oscillation
            frequency  as a function of the oscillation frequency
            ($\nu=\omega/2\pi$) for degenerate pairs $(\ell_1,\ell_2)=(0,2)$
	    corresponding to the fundamental radial mode (left panel) and
	    the first overtone (right panel). Circles and stars represent the
	    coupling strength coefficients for the radial modes and their
            corresponding quadrupole coupled modes, respectively. Those 
	    calculated for the selected models with $\Omega\sim41\,\kms$ are 
	    represented with filled symbols, and those calculated for 
	    the selected models with $\Omega\sim20\,\kms$ are represented by
            empty symbols.}
   \label{fig:Cab_p0p1}
 \end{center}
\end{figure*}
As shown in SGM06, the impact of near-degeneracy on the oscillation frequencies 
for $\ell=(0,2)$ degenerate pairs is quite significant for $p$ modes.
For such coupled pairs, only second-order coefficients are kept, thus we 
define the coefficient (SGM06, Eq.~57):
\eqn{C^{(2)}_{jk}\equiv \wdt-\omega_2\pm\frac{\delta \omega_0}{2}
\label{eq:defCij_12b}}
for near-degenerate modes with $\ell_j=\ell_{k}\pm2$, which represents 
the second-order frequency corrections for near-degeneracy, and $\omega_2$ is
the second-order frequency corrections free from near-degeneracy effects. 
For the sake of brevity, in the following, this coefficient is called
\emph{coupling strength}. We recall that the selection rules 
for second-order, near-degenerate modes restrict the coupling to modes with
$\Delta\,m=0$, i.e. to modes with the same azimuthal order. This means that 
radial modes can be coupled only with $(\ell,m)=(2,0)$ modes, which are 
expected to be less affected by near-degeneracy than for $m\neq0$ coupled 
modes (see Sect.~5.5 in SGM06 for more details).

In SGM06 the second-order near-degenerate coefficients 
$C_{jk}^{(2)}$ for $(\ell,m)=(2,0)$ coupled pairs
were studied for a typical \ds\ stellar model of $1.8\,\msol$, 
with solar metallicity. It was found that the coupling strength
increases with the rotational velocity. In the case 
of the fundamental radial mode and the first overtone, which are found 
typically to be in the range of 50-$200\,\muHz$ for \dss, the coupling strength 
$|C_{jk}^{(2)}|$ can vary from 0.05 to 0.13, approximately.

As shown in \paperI, the difference in period ratios increases with the
rotational velocity for a given metallicity. It is therefore necessary to study
the effect of coupling strength on the fundamental radial mode and the first
overtone when the rotational velocity and metallicity are varied. 
For this purpose we use the models of Table~\ref{tab:modelcarac}. 
In Fig.~\ref{fig:Cab_p0p1}, the normalised coupling strength
$C_{jk}^{(2)}/\omega_j$ for the fundamental radial mode $\omega_{0,\Omega}$ 
(left panel) and the first overtone $\omega_{1,\Omega}$ (right panel), together 
with the coupling strength coefficients of their corresponding coupled quadrupole 
pairs,  are displayed for the selected models. The normalisation of the coefficients 
to the oscillation frequency is necessary to remove contamination due to the effects 
of evolution. Notice that the behaviour of the
$C_{jk}^{(2)}/\omega_j$ for both the fundamental radial mode (left panel) and
the first overtone (right panel) is similar for the two rotational velocities.
The coupling strength is proportional to the rotational velocity, so logically
we expect larger effects for the fastest models. However, the
selected rotational velocities are too low to show significant differences. 
Moreover, the coupling strength of the first radial overtone is 
independent of the metallicity, showing values of $|C_{jk}^{(2)}/\omega_j|\lesssim0.03$ 
for all the selected models. In contrast, the coupling coefficient of the
fundamental radial mode shows a clear dependence upon the metallicity, showing 
absolute values increasing with the metallicity that range from roughly 0.05 to 
0.15. Such values imply variations from 0.005 to $0.1\,\muHz$ in the oscillation
frequencies, which represent a non-negligible percentage of the total effect 
of rotation on the oscillation frequencies. We suspect that, as for
the contamination coefficients, the influence of the quadrupole coupled modes is
mainly the responsible for such different sensitivities
of the $C_{jk}^{(2)}/\omega_j$ to the metallicity, although other explicit 
dependencies in the coupling terms, viz, distribution of
density and metallicity in the stellar interior, may also play an important
role (work in progress).

Furthermore, the effect of near-degeneracy on the oscillation frequencies also
depends on the closeness of the mode frequencies (the closer the mode
frequencies the larger the coupling strength). Due to this dependence, the coupling 
strength coefficients are sensitive to the evolutionary stage of the star
through the \emph{avoided-crossing} phenomenon (this is discussed later 
in Sect.~\ref{sec:impact_Z}). In \paperI\ (Fig.~3), it was shown that the first
radial overtone mode is, in general, more affected by rotation 
effects (without taking near-degeneracy effects into account) than the
fundamental radial mode. On the other hand, when near-degeneracy effects are 
considered, such a behaviour can be reversed, like in Fig.~\ref{fig:Cab_p0p1}. 

This different behaviour is responsible for the wriggles shown in 
Fig.~\ref{fig:w0w1_deg} by both the fundamental radial mode and the first overtone 
during the evolution (expressed as the logarithm of the period, in days, of the 
fundamental radial mode). These wriggles, which are visible in 
both modes, are then transfered to the period ratios. In Fig.~\ref{fig:pd_illust} 
we show how, when near-degeneracy is considered, wriggles modify the 
characteristic smoothed behaviour of RPD. In the case depicted here, wriggles
change the period ratios up to 0.02. 

This implies a significant increase of the uncertainty in mass and metallicity 
determinations using RPD. Nevertheless, for the rotational velocities
considered here, the wriggles remain in the range of period ratios typically
found for Pop.~I stars, [0.772, 0.776], which reduces possible confusion with other
radial mode period ratios, and helps to disentangle period ratios of nonradial
modes.
\section{The $\ratiorotdeg$ period ratios and the metallicity/mass determinations
\label{sec:impact_Z}}

For the sake of clarity, $\ratiorotdeg$ are used when referring to period ratios
corrected for near-degeneracy, in contrast to those used in \paperI, $\ratiorot$, 
which does not include them.
\begin{figure*}[t]
 \begin{center}  
   \includegraphics[width=8.8cm]{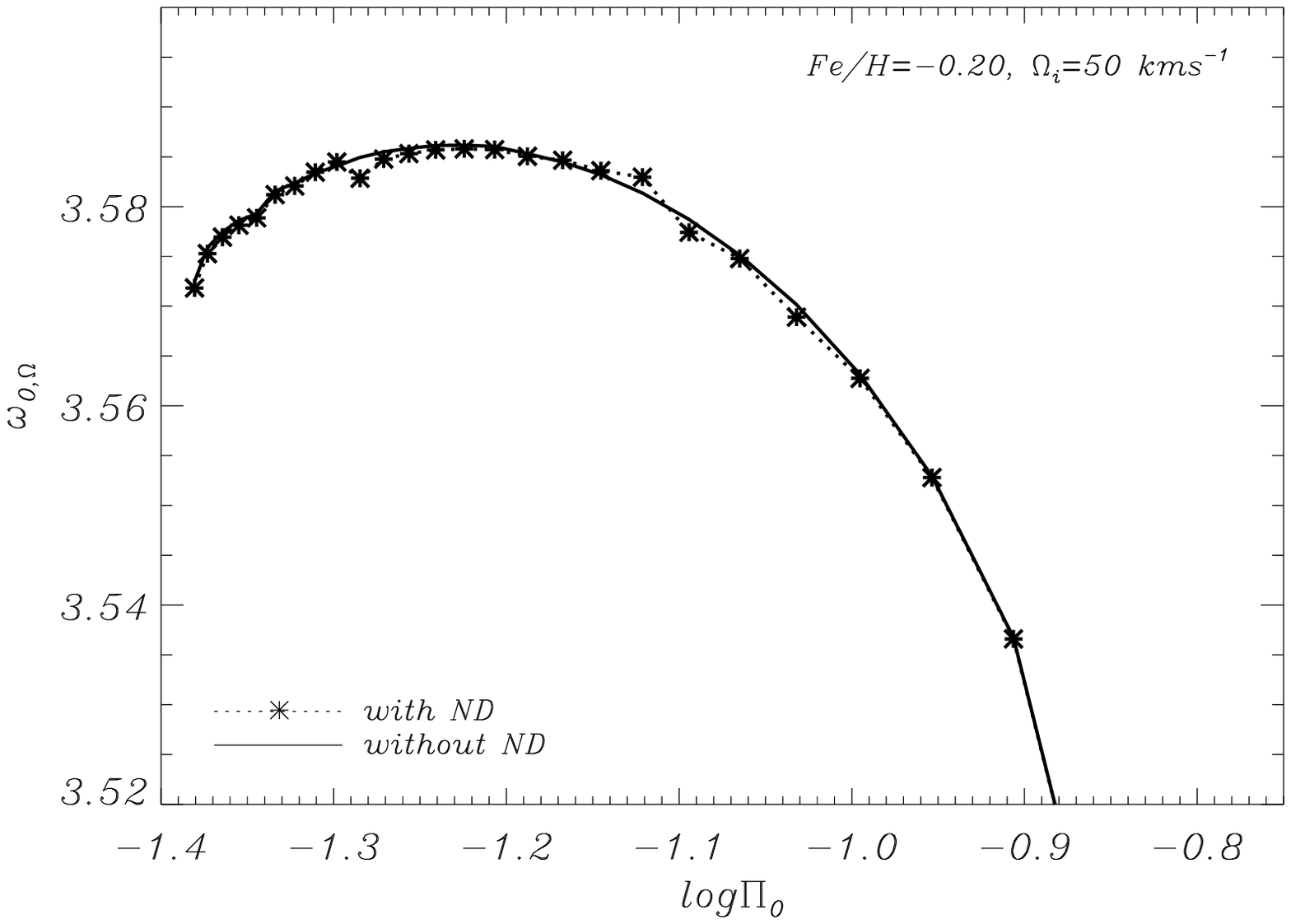}\hspace{-0.40cm}
   \includegraphics[width=8.8cm]{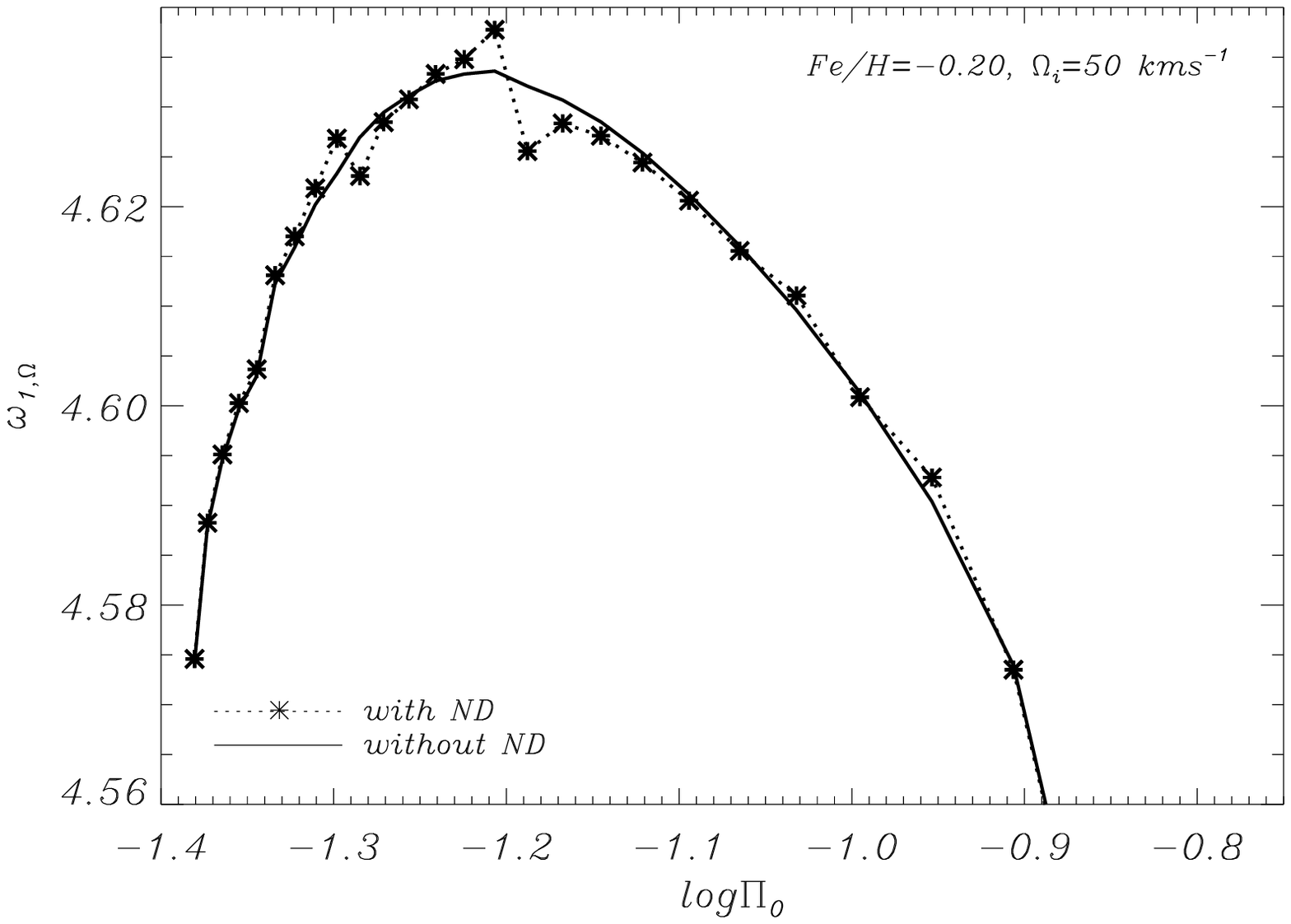}   
   \caption{Fundamental radial mode $\omega_{0,\Omega}$
           (left panel) and the first radial overtone $\omega_{1,\Omega}$ 
	   (right panel) normalised dimensionless frequencies (to $\sqrt{GM/R^3}$) 
	   as a function of the logarithm of the fundamental radial mode period (in $\mathrm{d}$).  
	   Tracks correspond to $1.8\,\msol$ models  with [Fe/H]=-0.20 and an 
	   initial rotational velocity of $\Omega_i=50\,\kms$. Dotted lines
	   with asterisks and solid lines represent frequencies corrected and 
	   non-corrected for near-degeneracy effects, respectively.}
   \label{fig:w0w1_deg}
 \end{center}
\end{figure*}
In order to determine how near-degeneracy affects metallicity determinations
using RPD, we proceed as in \paperI\ and select, from our model grid, several
evolutionary tracks with different metallicities:
$\mathrm{[Fe/H]}=0,-0.1,-0.2,-0.3,-0.35,-0.50$, and $-1.00\,\mathrm{dex}$,
two different initial rotational velocities $\Oi=25, 50$, and a 
fixed mass of $1.8\,\msol$. For each model we then computed the corresponding
$\ratiorotdeg$ period ratio. In Fig.~\ref{fig:rot_PD} we show RPD displaying
such period ratios, from top to bottom, for tracks computed for $\Oi=25$ to $50\,\kms$, 
respectively. As expected, the inclusion of the near-degeneracy effects in
the frequency corrections does not modify the general behaviour of the period
ratios with the metallicity, i.e. $\ratiorotdeg$ increases with increasing rotational 
velocities (left panels). As discussed in \paperI, this is equivalent to
a decrease in the metallicity in standard PD. However, the presence of wriggles may inverse
this situation in the regions where the curves cross each other.   
\begin{figure}
 \begin{center}
   \includegraphics[width=8.8cm]{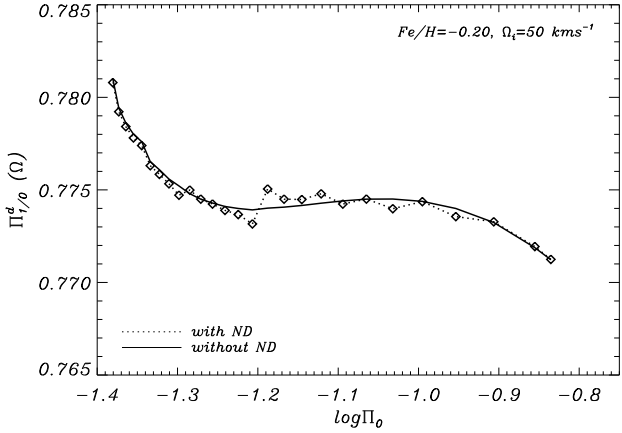}   
   \caption{RPD ($\Pi_0$ in $\mathrm{d}$) illustrating the effect of near-degeneracy
            on the period ratios. Tracks correspond to $1.8\,\msol$ models with 
	    [Fe/H]=-0.20. Wriggles are clearly visible when near-degeneracy (ND)
	    is taken into account.}
   \label{fig:pd_illust}
 \end{center}
\end{figure}
In order to compare quantitatively the evolution of $\ratiorotdeg$, with respect
to classical non-rotating PD, we use the period ratio differences defined as 
$$\dratiod(\Omega,\mathrm{[Fe/H]})=\Big[\ratiorotdeg-\ratio\Big]_{\mathrm{[Fe/H]}}$$
For the lowest rotational velocities ($\Oi\leq25\,\kms$), the size of wriggles
is almost negligible (Fig.~\ref{fig:rot_PD}, top panels), except for certain 
models; thus the discussion on the effect of rotation on period ratios 
given in \paperI\ remains valid, and the effect of near-degeneracy in
terms of period ratios differences is $\dratiod\sim2-3\cdot10^{-3}$.
For higher rotational velocities (see Fig.~\ref{fig:rot_PD}, bottom 
panels), wriggles are large which imply period ratios differences of 
$\dratiod\sim10^{-3}$ for some models, and close to $10^{-2}$ for others.  
The largest differences are found for main-sequence models, but
significant differences are also observed for more evolved evolutionary stages.

Let us now analyse the effect of near-degeneracy on RPD in terms of metallicity.
For shortness, we adopt the same nomenclature used in \paperI\ to identify the
different tracks as a function of the metallicity and rotational 
velocity. For instance, we use $[-0.1]_{25}$ when referring to models  
with $\Oi=25\,\kms$ and $\mathrm{[Fe/H]}=-0.1$. For the lowest initial
rotational velocity considered, $\Oi=25\,\kms$, the results are similar to those obtained
without considering near-degeneracy effects. This means that, even considering
the wriggles in the period ratios, the \emph{confusion} between tracks remains 
around 0.10-15\,dex. More important is the case of models with
$\Oi=50\,\kms$, for which large wriggles are responsible for significant
uncertainties in metallicity determinations. In particular, rotating tracks 
$\mathrm{[0.00]}_{50}$ may be confused with $[-0.35]_{0}$ tracks, and 
$\mathrm{[-0.10]}_{50}$ may be confused with non-rotating 
$\mathrm{[-0.50]}_{0}$ tracks. This implies that the uncertainty of metallicity
determinations may reach 0.50\,dex, which is critical for
Pop.~I HADS. 
\begin{figure*}
 \begin{center}
   \includegraphics[width=9cm]{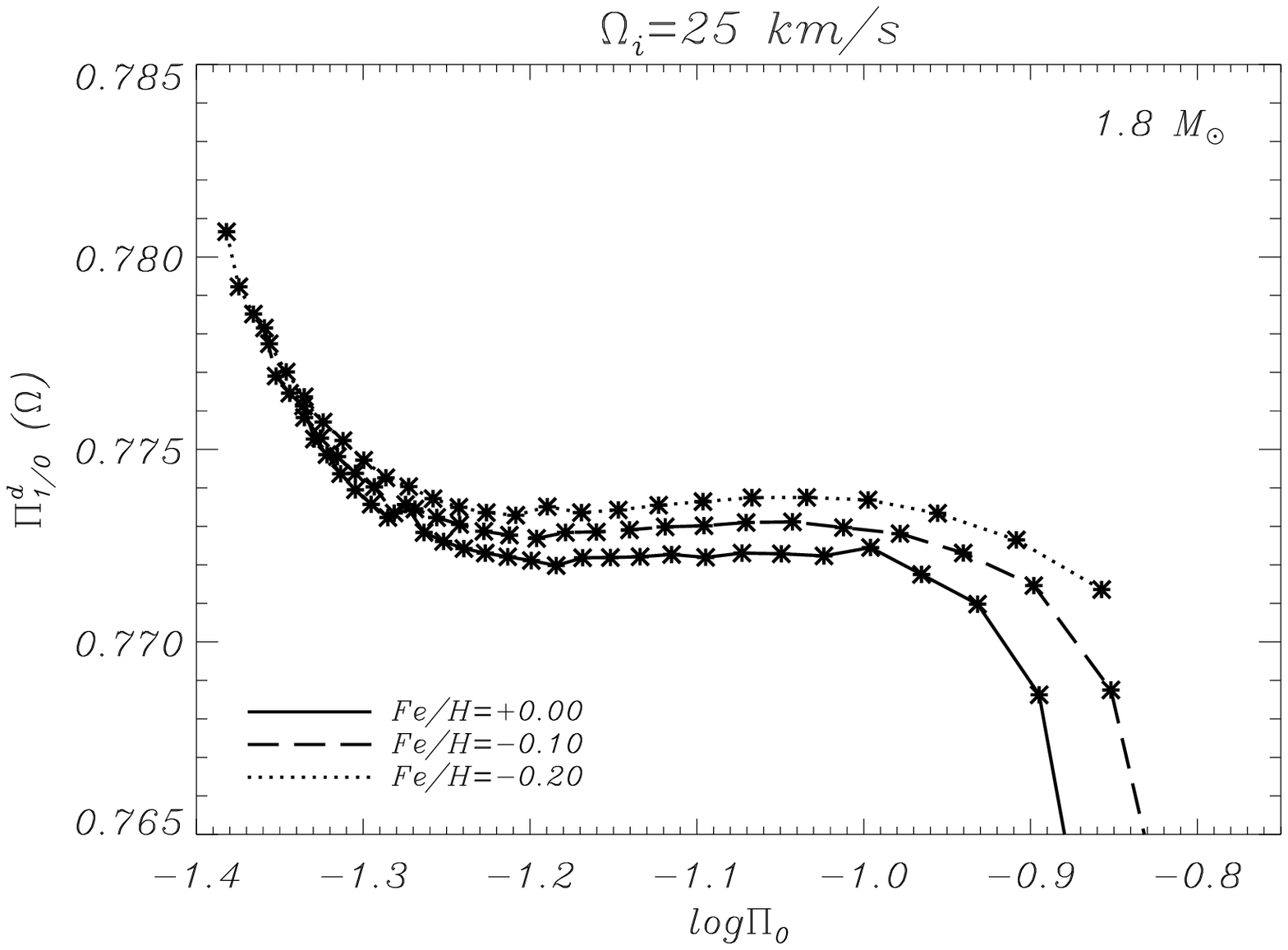}\hspace{-0.40cm}  
   \includegraphics[width=9cm]{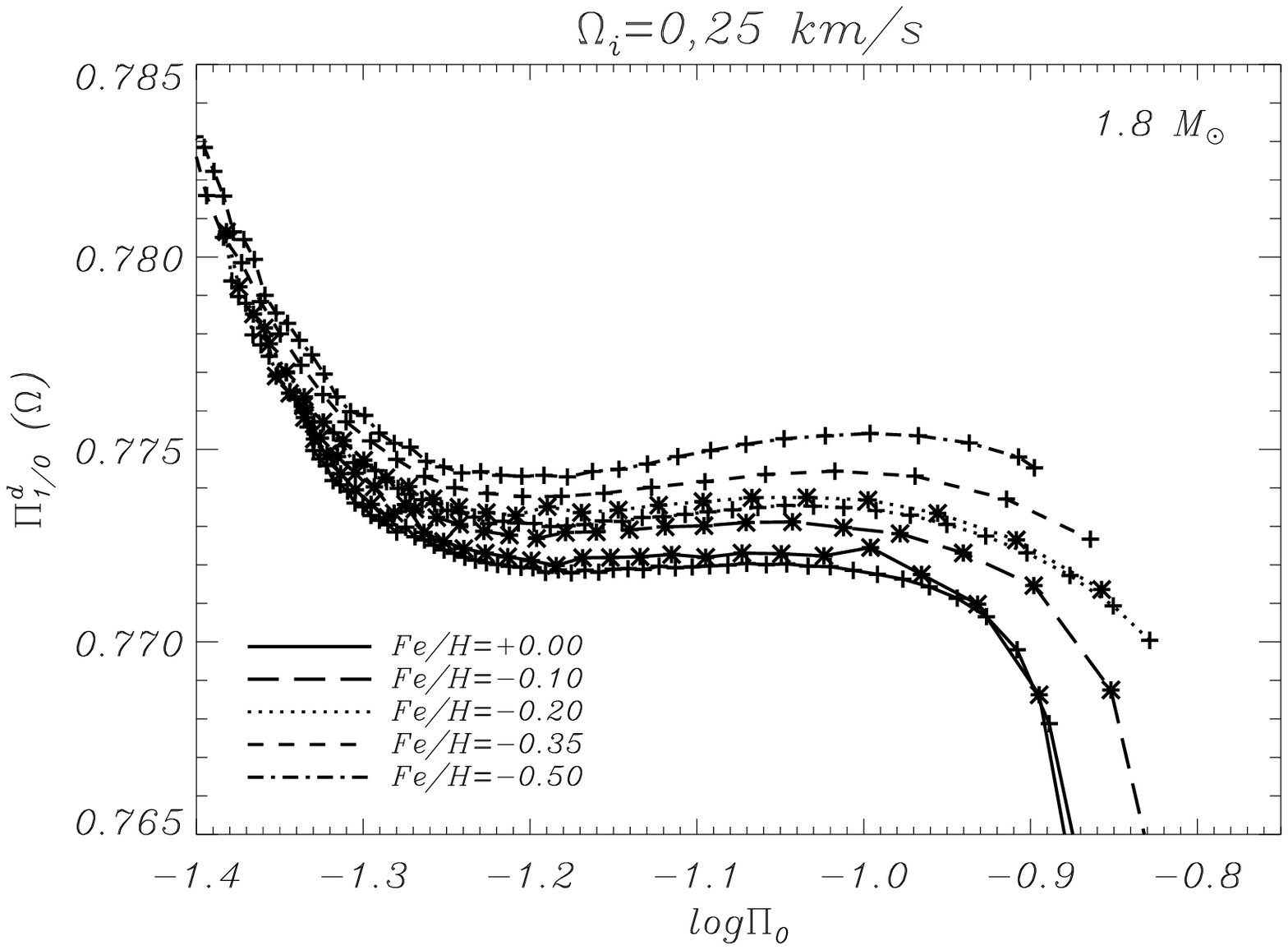}\hspace{-0.40cm} 
   \includegraphics[width=9cm]{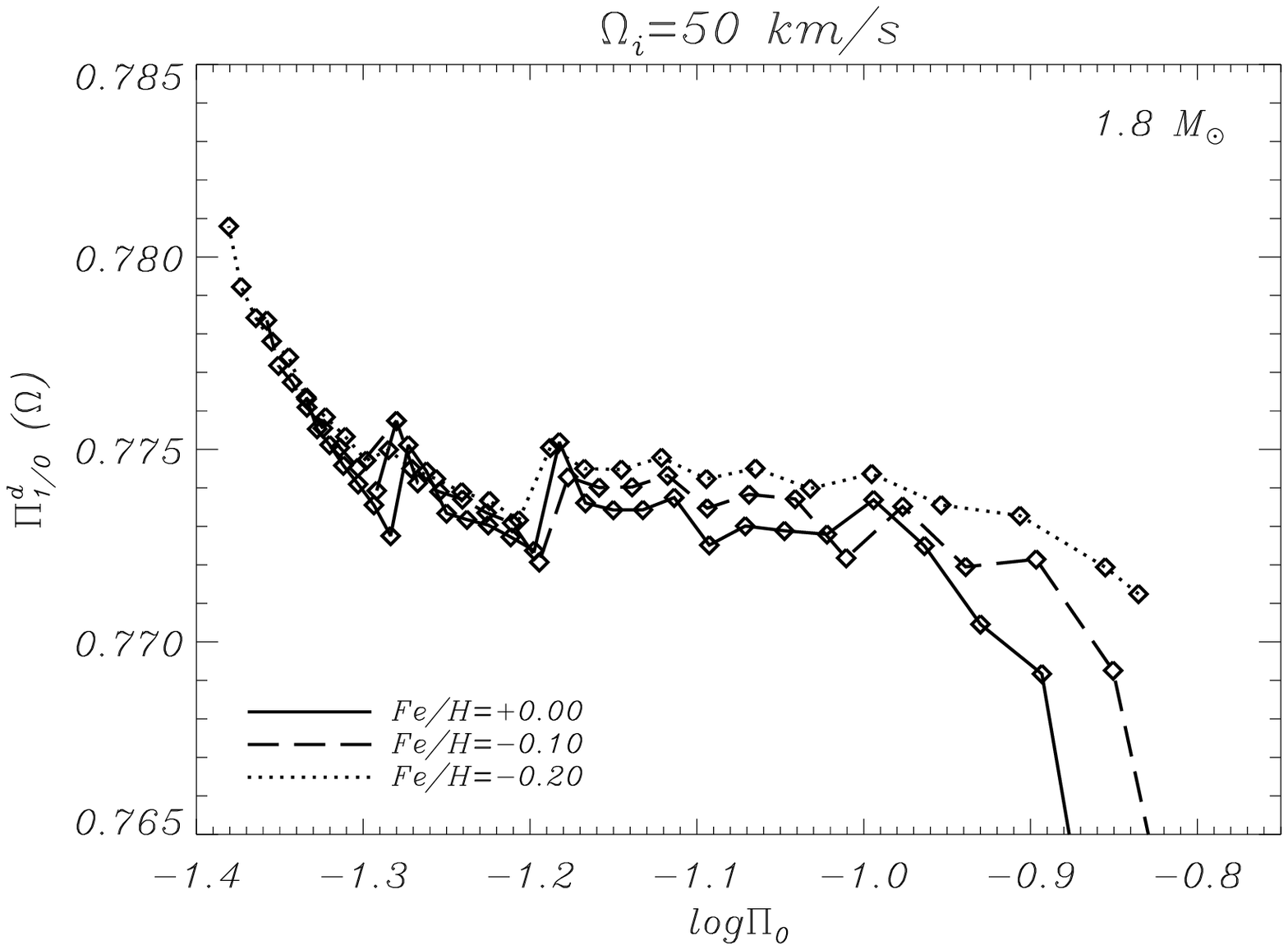}\hspace{-0.40cm}
   \includegraphics[width=9cm]{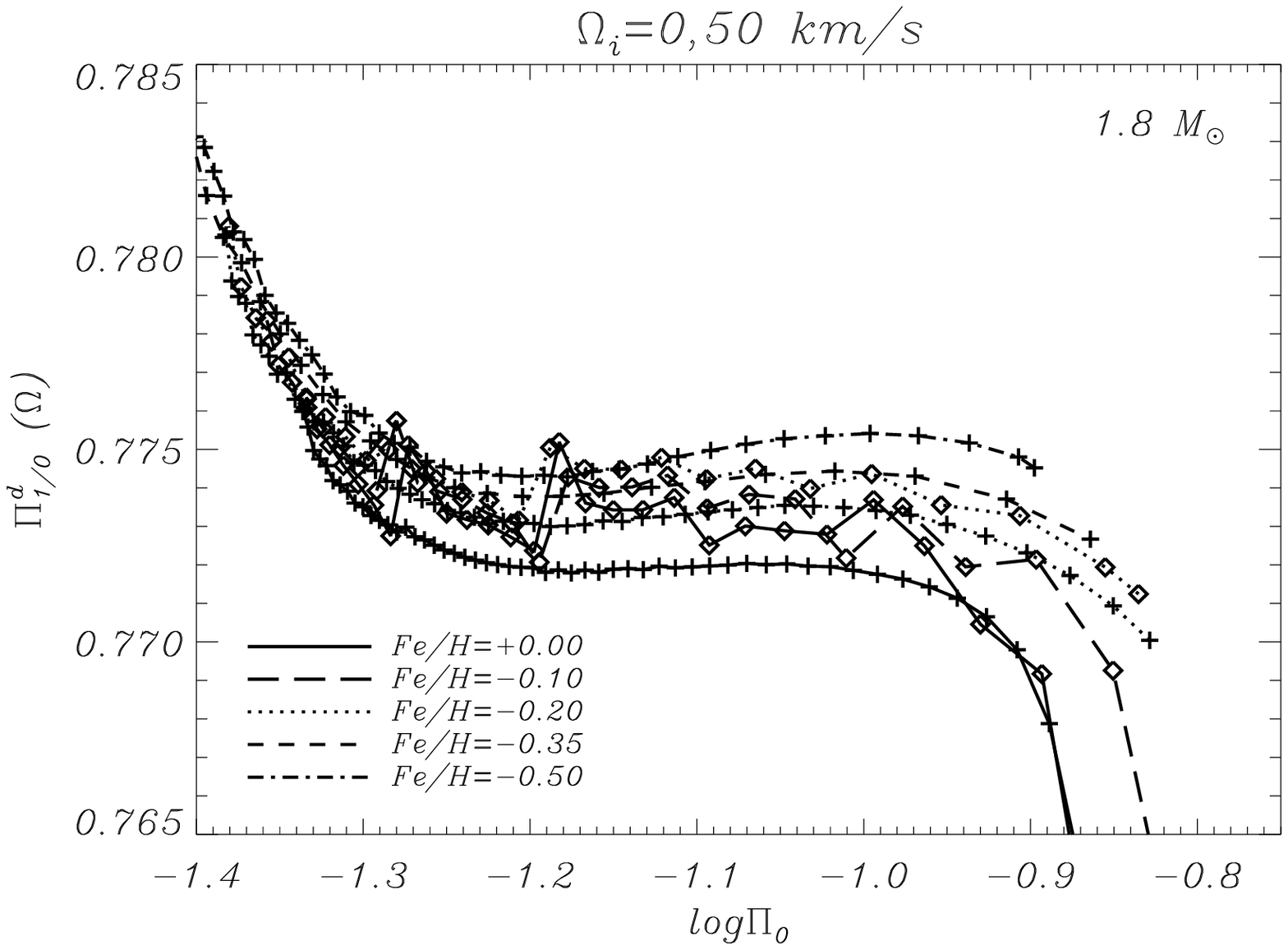}              
   \caption{Theoretical RPD including near-degeneracy effects on frequencies. 
            The $\ratiorotdeg$ period ratios have been computed in the manner 
	    described in Sect.~\ref{sec:modelling} for a set
	    of evolutionary $1.8\,\msol$ tracks obtained for different
	    metallicities. Tracks for two initial rotational velocities are 
	    considered: 25 and $50\,\kms$ (from top to bottom). Left panels
	    show only rotating models. Right panels show the comparison of
	    rotating and non-rotating tracks (classic PD). For convenience, the 
	    following symbols are used: crosses, representing non-rotating models;
	    asterisks, representing models evolved with $\Oi=25\,\kms$;
	    diamonds and those evolved with $\Oi=50\,\kms$.}
   \label{fig:rot_PD}
 \end{center} 
\end{figure*}
Analysis of the period ratios in terms of mass yields similar results to those
obtained in \paperI, i.e. $\ratiorotdeg$ increases with increase in
mass for any rotating track. In Fig.~\ref{fig:rot_PD_M}, we compare
rotating and non-rotating tracks for different masses. In particular,
it can be seen that the period ratios for $1.8\,\msol$ models with
$\Oi=50\,\kms$ and [Fe/H]=--0.20 can be confused with $2.30\,\msol$ 
non-rotating models with solar metallicity. This yields an uncertainty
of $0.5\,\msol$, which cannot be neglected in any mass determination
through PD. A detailed analysis of the uncertainty on period ratios
caused by rotation effects (including near-degeneracy), as well as by the mass 
and other physical parameters (namely the overshooting and mixing-length 
parameter), is currently in preparation.

In general, near-degeneracy effects complicate the already confusing scenario 
stated in \paperI, in which near-degeneracy was not considered. Nevertheless,
such an apparent worsening of the situation, may in fact be a source of 
information, provided that wriggles in PD are not placed randomly. Indeed,
rotational coupling is sensitive to the so-called avoided-crossing phenomenon,
i.e. the well-known phenomenon occurring when the convective core recedes during 
the main sequence evolution. The relation between near-degeneracy and the 
avoided-crossing phenomenon comes from the nature of the near-degeneracy itself,
which takes place  when two or more modes close in frequency fulfil certain selection 
rules (\dg, \soufi, SGM06). These selection rules restrict
the mode coupling to modes having the same azimuthal order $m$ and $\Delta\ell=0, \pm2$.
The case treated here concerns $\ell=0$ and ($\ell=2, m=0$) modes, which are
naturally close in frequency, and then systematically coupled. This can be easily seen
in Fig.~\ref{fig:n}, which shows the evolution of the radial order values of the quadrupole pairs 
($n_{0,\mathrm{c}}$ and $n_{1,\mathrm{c}}$) with which the fundamental radial mode and
the first overtone are coupled. During the evolution, the pairs ($n_{0,c}, n_{1,c}$) 
decrease their values following a given pattern, which is clearly modulated by the
aforementioned avoided-crossing phenomenon. For the radial modes, such a modulation is different
from the modulation occurring in the corresponding quadrupole coupled modes. That is,
the \emph{step} features shown in Fig.~\ref{fig:n} occurs for $n_{0,c}$ at different 
evolutionary stage than for $n_{1,c}$, which causes the wriggles in the period ratios. 
This can be verified by comparing Fig.~\ref{fig:n} with Fig.~\ref{fig:w0w1_deg} for modes 
around $\log\Pi_0\sim[-1.3,-1.2]$.
%
Interestingly, the results shown in Fig.~\ref{fig:n} are nearly
independent of the metallicity and the rotational velocity\footnote{In the
range of rotational velocities considered in this work.}, and present some slight
dependence upon the mass. This may be helpful to identify the quadrupole modes
coupled with the radial modes, together with other mode identification
techniques. For such cases, with accurate determinations of metallicity
provided, analysis of this phenomenon may also be able to help in constraining
the mass of the models.
\begin{figure}[t]
   \includegraphics[width=8.8cm]{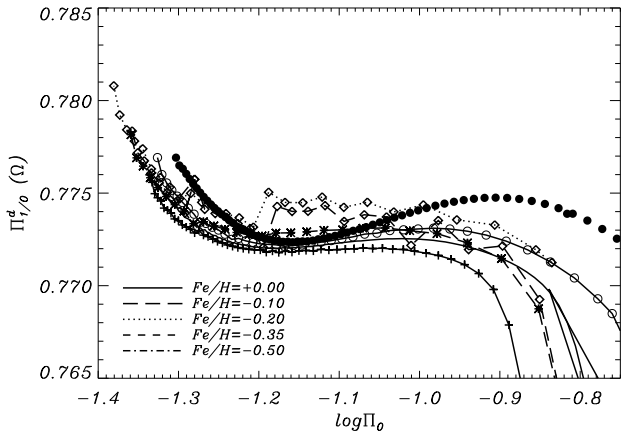}   
   \caption{RPD ($\Pi_0$ in $\mathrm{d}$) illustrating the effect of
            considering different masses. Symbols have the same
	    meaning of those from Fig.~\ref{fig:rot_PD}, except for
	    empty and filled circles, which correspond to $2.00$ and 
	    $2.30\,\msol$ non-rotating models and solar metallicity, 
	    respectively.}
   \label{fig:rot_PD_M}
\end{figure}

\section{Conclusions\label{sec:conclusions}}

   The effect of near-degeneracy in RPD has been 
   analysed.  Following \paperI, 
   detailed seismic models have been computed considering rotation effects on
   both equilibrium models and on adiabatic oscillation frequencies. 
   A grid of $1.8\,\msol$ stellar equilibrium models, similar to
   the one used in \paperI\ (completed with [Fe/H]=-0.30 models), has been 
   used. From these models, the corresponding oscillation eigenfrequencies 
   corrected for the effect of rotation up to the second order, including 
   near-degeneracy effects, have been computed. 
   
   Near-degeneracy effects have been divided into two components: the effect
   of mode \emph{contamination} and the coupling strength. 
   The former effect concerns the relative spherical harmonic amplitude of 
   degenerate modes. Analysis of such relative amplitudes (contamination 
   coefficients) reveals that the identity of the degenerate fundamental radial 
   mode remains almost unaltered, while the degenerate first overtone presents a mixed 
   radial/quadrupole identity.
   This must then be taken into account when performing $\ell$-diagnostics using
   amplitude ratio/phase difference diagrams from multicolour photometry.
   Indeed, analysis of mode contamination enhances such diagnostic diagrams
   since not only may it help to discriminate the fundamental radial mode
   from the first overtone, but also to identify the quadrupole modes coupled
   with them (if they are present in the observed spectrum). 

   The effect of near-degeneracy on the oscillation frequencies
   (coupling strength) has also been studied. Although this effect increases
   for increasing rotational velocities, for the rotational velocities 
   considered here, similar values of the couple strength coefficients are 
   found when changing the rotational velocity of the models.
   In terms of frequency variations, the coupling strength analysis 
   suggests that near-degeneracy may modify the frequency of the fundamental
   radial mode and the first overtone up to $0.3\,\muHz$, which represents 
   a non-negligible percentage of the total effect of rotation on the oscillation 
   frequencies. Moreover, near-degeneracy affects the frequency
   of the fundamental radial mode and the first overtone differently. In particular,     
   the largest effect on the frequencies is found alternatively for the fundamental 
   radial mode and the first overtone, depending on the evolutionary stage.  
   This alternation is responsible for the wriggles shown in the 
   fundamental-to-first overtone period ratios when near-degeneracy effects are 
   taken into account. 
   Analysis of these wriggles in the tracks, which are larger for increasing 
   rotational velocities, reveals that they become significant (in the context
   of RPD) and degrade substantially the accuracy of period ratios $\ratiorotdeg$,
   which can reach up $10^{-2}$. 
   In terms of metallicity, this implies 
   uncertainties reaching up to 0.50\,dex (for the largest rotational
   velocity considered), which is critical for
   Pop.~I HADS. Furthermore, near-degeneracy effects also increase the uncertainty in mass
   determination using RPD, which can reach $0.5\,\msol$ for certain
   models. In general, for stellar rotational velocities larger
   than $15-20\,\kms$, this new scenario would thus invalidate, a priori, the
   PD as diagnostic diagrams. 
   However, it is found that wriggles in period ratios are not located randomly in PD.
   They depend on the frequency evolution of the quadrupole modes coupled
   with the radial ones, which are mainly modulated by the avoided-crossing
   phenomenon. Moreover, it is found that such a behaviour 
   is dependent on the mass of the star, but is nearly independent  
   of the rotational velocity and  metallicity. These results
   provide clues for the mode identification of
   the fundamental radial mode, the first overtone, and their 
   corresponding quadrupole coupled modes, using only \emph{white} light 
   photometry. If additional information from multicolour 
   photometry and/or spectroscopy is given, RPD may be a powerful
   asteroseismic technique not only for diagnostics on metallicity and/or mass,
   but also for the rotational velocity (and thereby the
   inclination angle). 
\begin{figure}[t]
 \begin{center}
   \includegraphics[width=8.8cm]{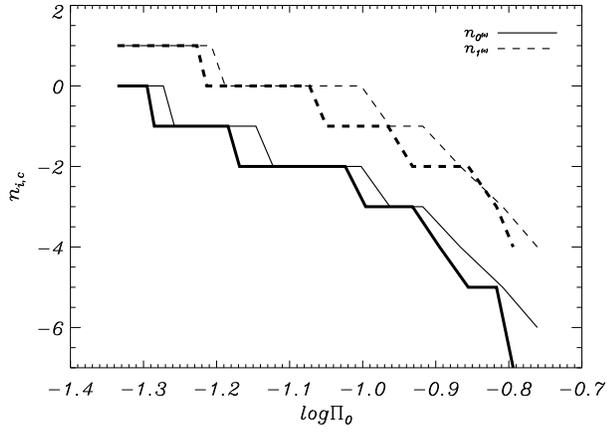}
   \caption{Radial order $n_{0,\mathrm{(c)}}$ and $n_{1,\mathrm{(c)}}$  of the 
            $\ell=2, m=0$ pairs respectively coupled with the fundamental radial 
	    mode and the first overtone, as a function of the logarithm of the 
	    fundamental radial mode period (in days). Thick lines correspond to 
	    models of $1.8\,\msol$, and thin lines to models with $2\,\msol$, 
	    both computed with $\Oi=25\,\kms$ and solar metallicity.}
   \label{fig:n}
 \end{center}
\end{figure}

\acknowledgements{This study would not have been possible without the financial
support of the Instituto de Astrof\'{\i}sica de Andaluc\'{\i}a (CSIC) by an I3P
contract financed by the European Social Fund and from the Spanish
"Plan Nacional del Espacio" under project ESP2004-03855-C03-01.}
\bibliography{/home/jcsuarez/Boulot/Latex/Util/References/ref-generale}

\begin{thebibliography}{13}
\expandafter\ifx\csname natexlab\endcsname\relax\def\natexlab#1{#1}\fi

\bibitem[{{Casas} {et~al.}(2006){Casas}, {Su{\'a}rez}, {Moya}, \&
  {Garrido}}]{Casas06}
{Casas}, R., {Su{\'a}rez}, J.~C., {Moya}, A., \& {Garrido}, R. 2006, \aap, 455,
  1019

\bibitem[{{Daszy{\' n}ska-Daszkiewicz} {et~al.}(2002){Daszy{\'
  n}ska-Daszkiewicz}, {Dziembowski}, {Pamyatnykh}, \& {Goupil}}]{Pagoda02}
{Daszy{\' n}ska-Daszkiewicz}, J., {Dziembowski}, W.~A., {Pamyatnykh}, A.~A., \&
  {Goupil}, M.-J. 2002, \aap, 392, 151

\bibitem[{{Dziembowski} \& {Goode}(1992)}]{DG92}
{Dziembowski}, W.~A. \& {Goode}, P.~R. 1992, \apj, 394, 670

\bibitem[{{Garrido}(2000)}]{Garrido00}
{Garrido}, R. 2000, in Delta Scuti and Related Wtars, Reference Handbook and
  Proceedings of the 6th Vienna Workshop in Astrophysics, held in Vienna,
  Austria, 4-7 August, 1999. ASP Conference Series, Vol. 210. Edited by Michel
  Breger and Michael Montgomery. (San Francisco: ASP) ISBN: 1-58381-041-2, 67

\bibitem[{{Lochard} {et~al.}(2007){Lochard}, {Su{\' a}rez}, \&
  {Goupil}}]{Lochard07}
{Lochard}, J., {Su{\' a}rez}, \& {Goupil}, M.~J. 2007, in preparation

\bibitem[{{Morel}(1997)}]{Morel97}
{Morel}, P. 1997, \aaps, 124, 597

\bibitem[{{Pamyatnykh}(2003)}]{Alosha03}
{Pamyatnykh}, A.~A. 2003, \apss, 284, 97

\bibitem[{{Reese} {et~al.}(2006){Reese}, {Ligni{\`e}res}, \&
  {Rieutord}}]{Reese06}
{Reese}, D., {Ligni{\`e}res}, F., \& {Rieutord}, M. 2006, \aap, 455, 621

\bibitem[{{Soufi} {et~al.}(1998){Soufi}, {Goupil}, \& {Dziembowski}}]{Soufi98}
{Soufi}, F., {Goupil}, M.~J., \& {Dziembowski}, W.~A. 1998, \aap, 334, 911

\bibitem[{{Su{\' a}rez}(2002)}]{SuaThesis}
{Su{\' a}rez}, J.~C. 2002, Ph.D.~Thesis, ISBN 84-689-3851-3, ID 02/PA07/7178

\bibitem[{{Su{\'a}rez} {et~al.}(2006{\natexlab{a}}){Su{\'a}rez}, {Garrido}, \&
  {Goupil}}]{Sua06pdrotPaperI}
{Su{\'a}rez}, J.~C., {Garrido}, R., \& {Goupil}, M.~J. 2006{\natexlab{a}},
  \aap, 447, 649~(Paper~I)

\bibitem[{{Su{\'a}rez} {et~al.}(2006{\natexlab{b}}){Su{\'a}rez}, {Goupil}, \&
  {Morel}}]{Sua06rotcelSGM06}
{Su{\'a}rez}, J.~C., {Goupil}, M.~J., \& {Morel}, P. 2006{\natexlab{b}}, \aap,
  449, 673~(SGM06)

\bibitem[{{Tran Minh} \& {Léon}(1995)}]{filou}
{Tran Minh}, F. \& {Léon}, L. 1995, in Roxburg, I.~W., Maxnou, J.~L., eds.,
  Physical Processes in Astrophysics. Springer Verlag, Berlin., 219

\end{thebibliography}
\bibliographystyle{aa}

 \end{document}